\documentclass[10pt,prd,aps,twocolumn,showpacs,showkeys
,superscriptaddress,groupedaddress,floatfix]{revtex4-1}
\usepackage{bm}
\usepackage{amsmath}
\usepackage{amsfonts}
\usepackage{amssymb}
\usepackage{graphicx}
\usepackage{graphics}
\usepackage[normalem]{ulem}
\usepackage{soul}
\usepackage[dvips]{color}
\usepackage[latin1]{inputenc}
\usepackage[english]{babel}
\usepackage{anysize}

\newcommand{\be}{\begin{equation}}
\newcommand{\ba}{\begin{eqnarray}}
\newcommand{\ea}{\end{eqnarray}}

\def\ee{\mbox{$\left(e,e^{\prime}\right)$\ }}
\def\eep{\mbox{$\left(e,e^{\prime}p\right)$\ }}

\begin{document}

\author{Andrea Meucci}
\author{Carlotta Giusti}
\affiliation{Dipartimento di Fisica,
Universit\`{a} degli Studi di Pavia and \\
INFN,
Sezione di Pavia, via A. Bassi 6, I-27100 Pavia, Italy}

\title{The relativistic Green's function model 
in charged-current quasielastic 
neutrino and antineutrino scattering at MINER$\nu$A kinematics}

\date{\today}

\begin{abstract}
The analysis of charged-current quasielastic neutrino and antineutrino-nucleus 
scattering cross sections 
requires relativistic theoretical descriptions 
also accounting for the role of final-state interactions. 
We compare the results of the relativistic Green's 
function model 
with the data recently published  by the MINER$\nu$A Collaboration. The model 
is able to describe both  MINER$\nu$A  and MiniBooNE data.
\end{abstract}


\maketitle

\section{Introduction}
\label{intro}

In the past decade several Collaborations have presented their results of 
neutrino oscillations 
\cite{superkam,superkam11,icarus13,sno,minos12,minos13,minos14,
t2k12,t2k13,t2k14,minibPhysRevLett13,minibPhysRevD12,PhysRevD.86.052009,
PhysRevLett.108.131801,PhysRevLett.108.171803,
An:2012bu,PhysRevLett.108.191802,PhysRevD.74.072003,PhysRevD.64.112007}
that aim at a precise determination of mass-squared splitting and 
mixing angles in $\nu_{\mu}$  disappearance
and $\nu_{e}$ appearance measurements. 
In addition, various experimental neutrino-nucleus differential cross sections  
have been published \cite{miniboone,miniboonenc,miniboone-ant,Nakajima:2010fp,argoneut,
miniboonenc-nubar,PhysRevD.87.092003,t2k14n} 
and other measurements are planned in the near future. 
The reduction of uncertainties in baseline neutrino oscillation experiments is mandatory 
to obtain a deeper understanding of neutrino physics.
In this spirit, since experiments are performed with 
detectors made of heavy nuclear targets, e.g, Carbon, Oxygen, or Argon,
a clear understanding of nuclear effects is extremely important for the analysis 
of data. The recent progress, the questions and challenges
in the physics of neutrino cross sections are  reviewed in \cite{Morfin:2012kn,nieves:2014}.

The MINER$\nu$A  Collaboration has recently measured
differential cross sections for neutrino and antineutrino
charged-current quasielastic (CCQE) scattering on a hydrocarbon target  
in an energy range between 1.5 and 10 GeV \cite{PhysRevLett.111.022502,
PhysRevLett.111.022501}.  A large fraction  of the
total reaction cross section at the GeV energy scale can be ascribed to 
CCQE reactions, defined in this case as containing no mesons in
the final state, that can therefore be viewed as a reference 
for neutrino oscillation experiments in this energy range. 

The first measurements of the CCQE
flux-averaged double-differential $\nu_{\mu} (\bar{\nu_{\mu}})$ cross
section on $^{12}$C in the few GeV region by the MiniBooNE 
Collaboration  \cite{miniboone,miniboone-ant} have raised  extensive
discussions. Indeed, the fact that the experimental cross sections are usually 
 underestimated by the
relativistic Fermi gas  model and by other more sophisticated models 
based on the impulse approximation (IA)
\cite{Benhar:2010nx,Benhar:2011wy,Butkevich:2010cr,Butkevich:2011fu,jusz10}, unless 
the nucleon axial mass is significantly enlarged, up to 
$M_A \sim 1.2 \div 1.4$  GeV/$c^2$, with 
respect to the world average value of 1.03 GeV/$c^2$ \cite{Bernard:2001rs,bodek08},
 have suggested that effects beyond the IA may play a significant role in 
 this energy domain  \cite{PhysRevC.79.034601,Leitner:2010kp,
PhysRevC.83.054616,FernandezMartinez2011477,AmaroAntSusa,Amaro:2011qb,
Martini:2010ex,Martini:2013sha,Nieves201390,Golan:2013jtj}.

Models developed for QE electron scattering~\cite{Boffi:1993gs,book} and 
able to successfully describe a wide number of experimental data can provide 
a useful tool to study neutrino-induced processes. 
In particular, a reliable description of the effects
of the final-state interactions (FSI) between the ejected nucleon
and the residual nucleus is very important for the comparison
with data. The important role of FSI  has been clearly stated for the exclusive 
$\eep$ reaction 
within the framework of the distorted-wave impulse approximation (DWIA),
where the use of a complex optical potential (OP) with its absorptive 
imaginary part produces a reduction of the calculated cross section that is
essential to reproduce the data. 
The imaginary part of the OP accounts for the fact that in the elastic 
nucleon-nucleus scattering, if other channels are open besides the elastic one, 
part of the incident flux is lost in the elastically scattered beam and goes 
to the inelastic channels which are open. In the exclusive \eep reaction 
only one channel contributes and it is correct to account for the 
flux lost in the selected channel.
In the case of the inclusive $\ee$ reaction, as well as of CCQE neutrino scattering, 
all elastic and inelastic channels contribute, the total flux is
redistributed in all the channels but must be conserved, and the use of the 
DWIA with an absorptive complex OP is conceptually wrong. Different
approaches have been adopted within the framework of the relativistic IA (RIA) 
to describe FSI in the inclusive QE electron and neutrino-nucleus scattering.

In the relativistic plane-wave impulse approximation (RPWIA) FSI are neglected.
The results of this simple approach are usually significantly 
different from the data and, in addition, they do not reproduce the behavior of
the phenomenological scaling
function extracted from QE longitudinal $\ee$ data \cite{Maieron:2003df,Caballero:2006wi}. 
In other approaches based on the RIA, FSI are included in the emitted 
nucleon state with real potentials, either retaining only the real part of the 
relativistic energy-dependent complex optical potential (rROP), or using 
distorted waves obtained with the same relativistic 
energy-independent potentials  considered in describing the initial nucleon 
state (RMF) \cite{Maieron:2003df,Caballero:2005sn,minerva-juan}. 

In the relativistic Green's function (RGF) model FSI are described in the 
inclusive scattering consistently with the exclusive scattering by the same 
complex OP, the components of the nuclear response are written in terms of
matrix elements of the same type as the DWIA ones of the
exclusive  $\eep$ process, but involve eigenfunctions of the
OP and of its Hermitian conjugate, where the opposite sign of the imaginary part  
gives in one case an absorption and in the other case a gain of strength.   
The imaginary part is therefore responsible for the redistribution of 
the flux in all the channels and in the sum over all the channels the total 
flux is conserved. 
The RGF model has been extensively tested against QE $\ee$ data over a wide 
energy range and for different target nuclei \cite{Meucci:2003uy,Meucci:2009nm,esotici2,
PhysRevC.89.034604}. The detailed description of the model can be
found in our previous papers \cite{Capuzzi:1991qd,Meucci:2003uy,Meucci:2003cv,Capuzzi:2004au,
Meucci:2005pk,Meucci:2009nm,Meucci:2011pi,Meucci:ant}.

Thes results of these different descriptions of FSI have been compared in~\cite{Meucci:2009nm} 
for the inclusive QE electron scattering, in \cite{Meucci:2011pi} for 
the CCQE neutrino scattering, and in \cite{Meucci:2011vd,PhysRevC.88.025502} 
with the CCQE and NCE MiniBooNE data. 
Electron scattering data and their related scaling
functions are successfully described by both
RMF and RGF models. In the case of MiniBooNE neutrino data, both models reproduce the 
shape of the experimental CCQE cross sections, but only the RGF gives cross sections 
of the same magnitude as the experimental ones without the need to 
increase the world average value of $M_A$~\cite{Meucci:2011vd,Meucci:ant}. 
The larger RGF cross sections are due to the 
translation to the inclusive strength of the overall effect of inelastic 
channels that are recovered in the model by the imaginary part of the
relativistic OP and that 
are not included in the RMF and in other models based on the IA. 

It has been pointed out in \cite{Megias2013170} that 
IA-based models are able to reproduce the higher energy data from the 
NOMAD experiment \cite{Lyubushkin:2008pe}. 
The kinematics of the MINER$\nu$A experiment is higher than MiniBooNE but lower than NOMAD
and can be useful to understand the role of nuclear structure,
many-body mechanisms, and reaction models in neutrino-nucleus scattering.
We note that the RMF model provides a good description of CCQE MINER$\nu$A data \cite{minerva-juan}.

In this paper we compare the neutrino and antineutrino CCQE MINER$\nu$A 
cross sections with the results of our RGF model.

 \section{Results } 
 \label{results}

\begin{figure}[b]
    \centering
        \includegraphics[scale=.42, bb=-12 12 570 497, clip]{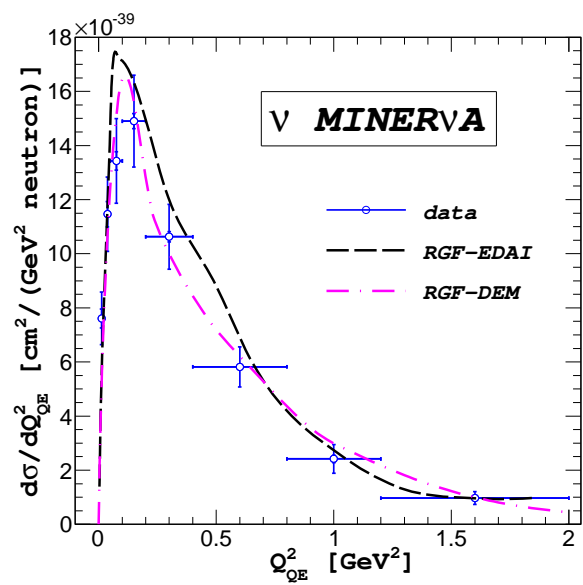}
     \caption{(Color online)
    CCQE flux-averaged $\nu - ^{12}$C  cross section per target nucleon as
    a function of $Q^2_{QE}$. 
    The data, with statistic and systematic errors, 
    are from MINER$\nu$A \cite{PhysRevLett.111.022502}.}
    \label{fig:minervanu}
\vspace*{-0.3 cm}
\end{figure}
\begin{figure}[t]
    \centering
         \includegraphics[scale=0.42, bb=-12 22 570 497, clip]{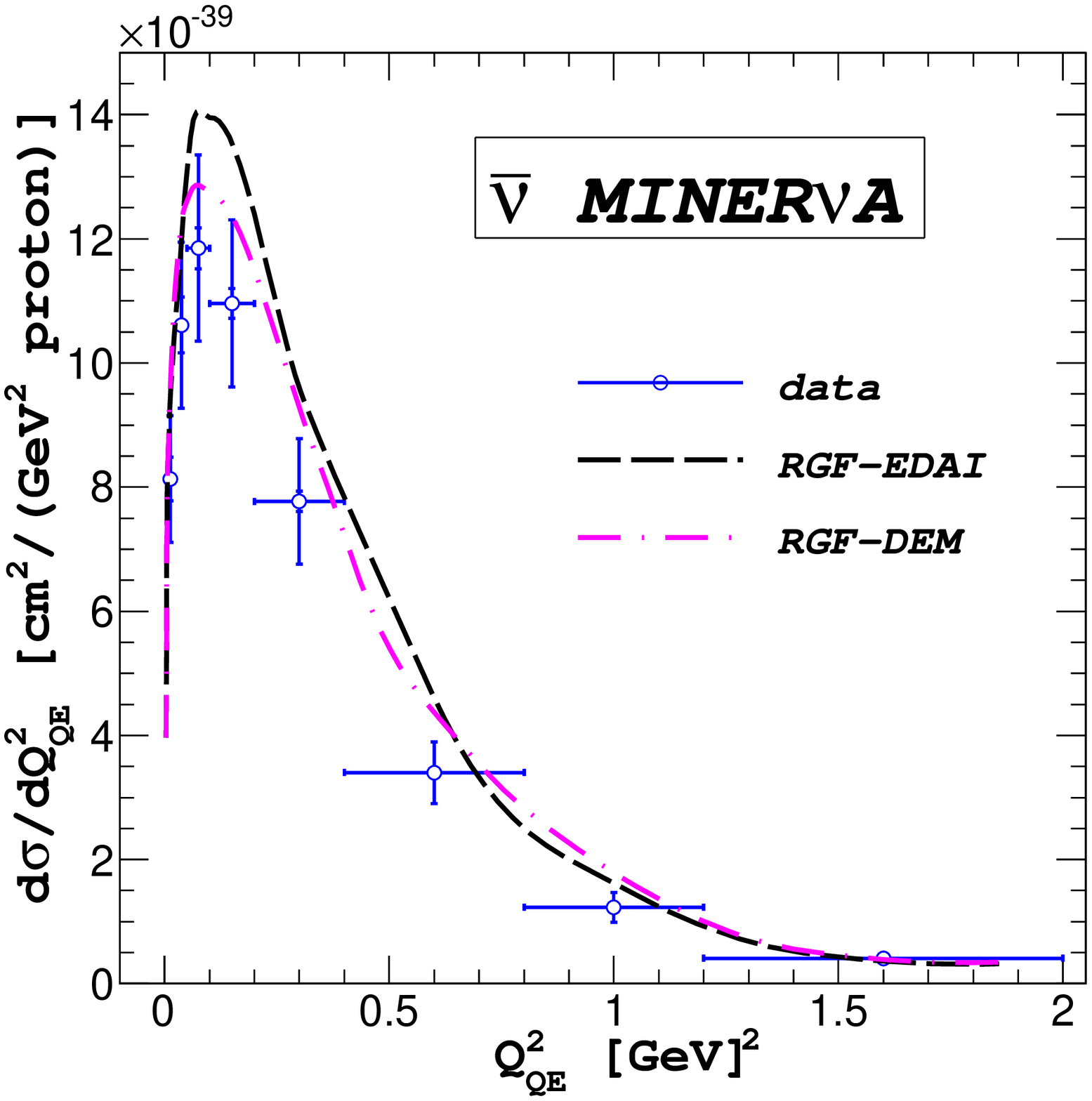}
     \caption{(Color online)
    CCQE flux-averaged $\bar{\nu} - ^{12}$C  cross section per target nucleon as
    a function of $Q^2_{QE}$. 
    The data, with statistic and systematic errors, 
     are from MINER$\nu$A \cite{PhysRevLett.111.022501}.}
    \label{fig:minervanubar}
\vspace*{-0.3 cm}
\end{figure}
\begin{figure}[th]
    \centering
\includegraphics[scale=0.42, bb=-12 2 570 447, clip]{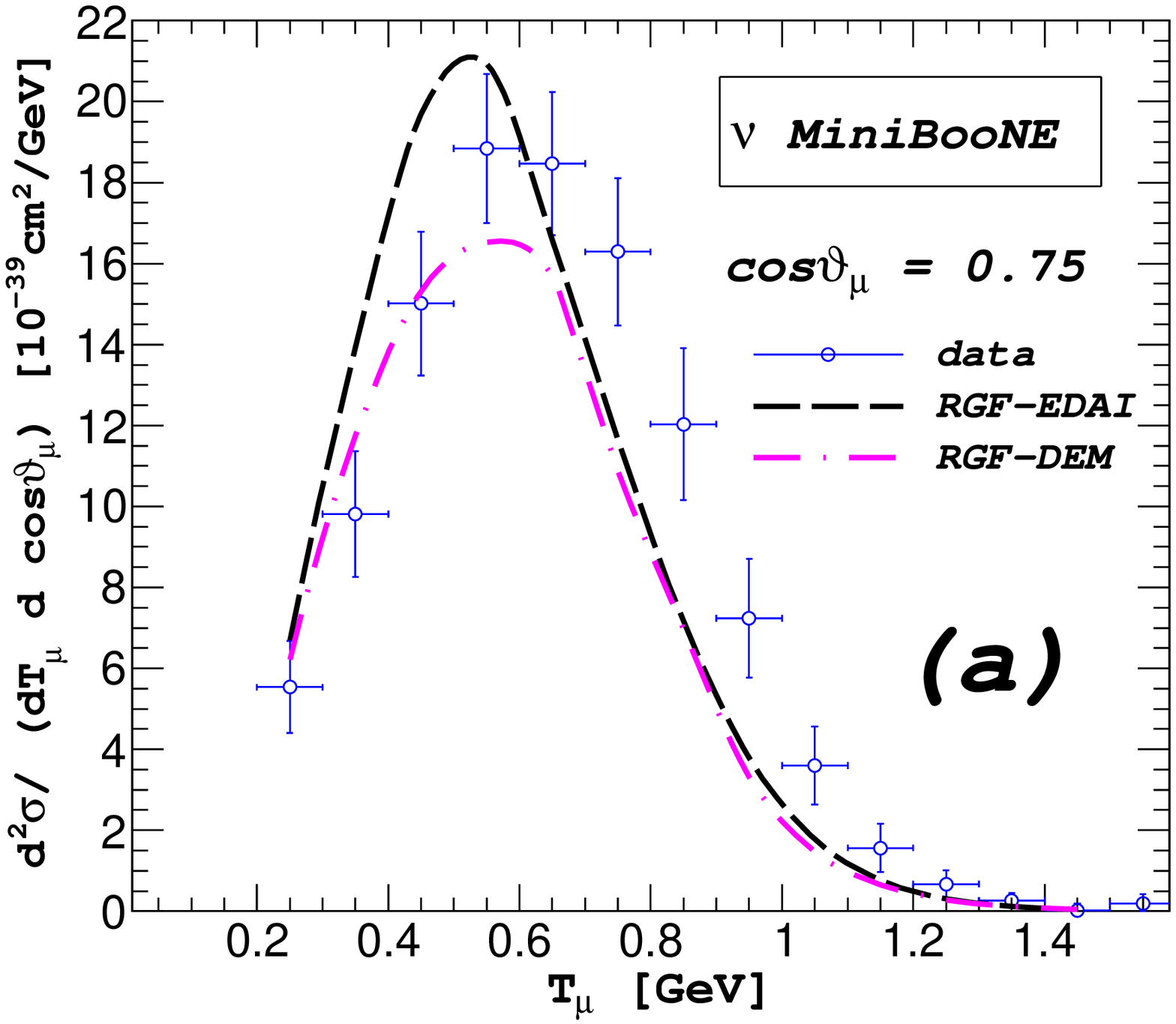}\\
\includegraphics[scale=0.42, bb=-12 2 570 447, clip]{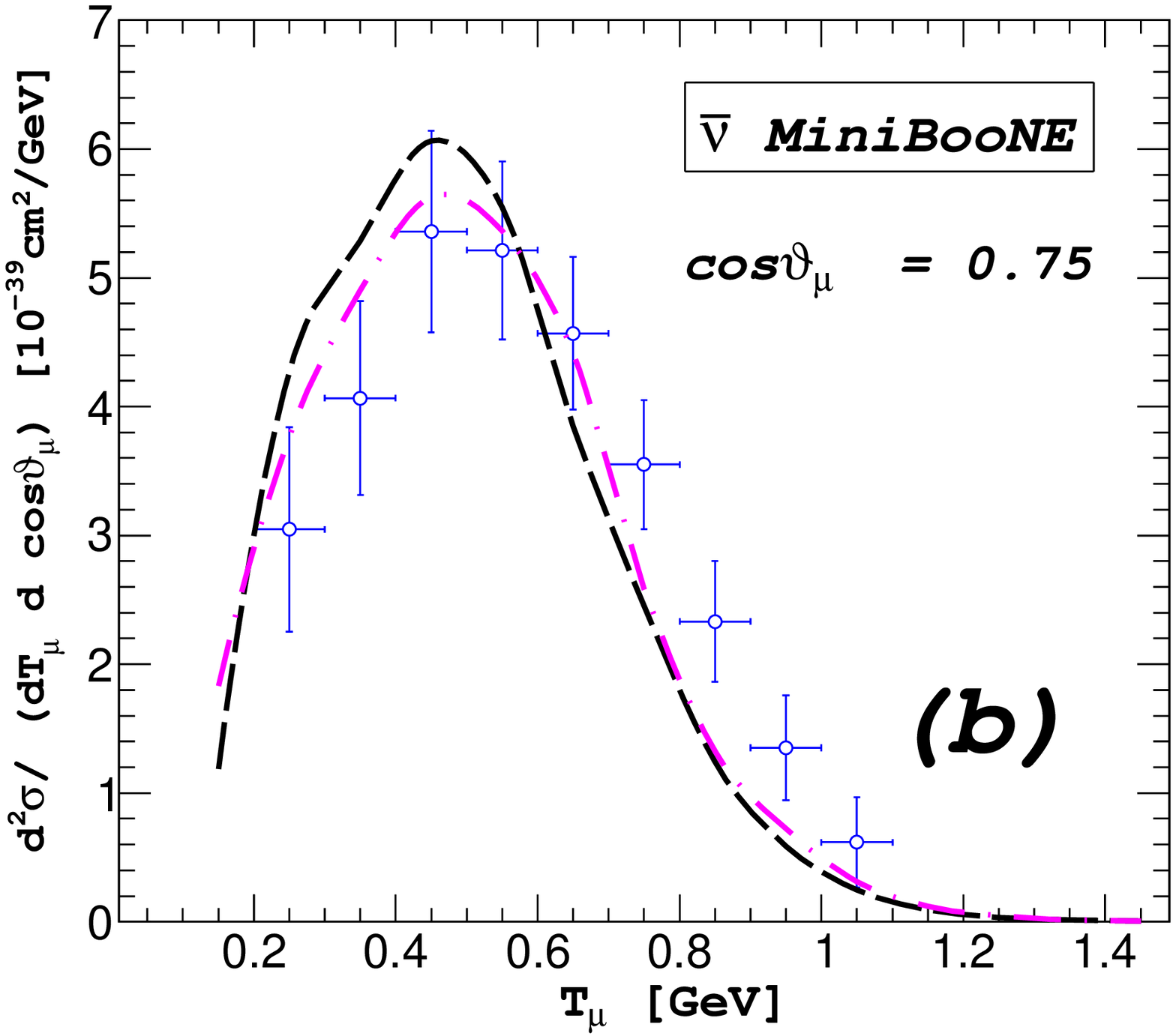}	
\caption{(color online) Flux-averaged double-differential cross section per 
	target nucleon for 
the CCQE neutrino [panel (a)] and antineutrino [panel (b)] reaction as 
a function of outgoing muon kinetic energy $T_{\mu}$ 
for the $\cos\vartheta_{\mu} = 0.75$ angular bin. 
 The data are from MiniBooNE \cite{miniboone,miniboone-ant}.
	   }
 \label{fig:miniboone_075}
\vspace*{-0.3 cm}
\end{figure}


In all the calculations presented in this work 
we have adopted the standard value for the nucleon axial mass 
$M_A = 1.03$ GeV$/c^2$. The bound nucleon states 
are taken as self-consistent Dirac-Hartree solutions derived 
within a relativistic mean field approach using a Lagrangian containing 
$\sigma$, $\omega$, and $\rho$ mesons 
\cite{Serot:1984ey,Rein:1989,Rign:1996,Lalazissis:1996rd,Serot:1997xg}.
A crucial ingredient of the RGF calculations is the relativistic OP. 
We have used two different parametrizations for the OP of $^{12}$C: the
Energy-Dependent and A-Independent EDAI
(where the $E$ represents the energy and the $A$ the atomic number) 
 OP of \cite{Cooper:1993nx}, 
and  the more recent Democratic (DEM)
phenomenological OP of \cite{Cooper:2009}. The
EDAI OP is a single-nucleus parametrization which is constructed
to better reproduce the elastic proton-$^{12}$C phenomenology, whereas 
 the DEM parametrization is a global parametrization, which
 depends on the atomic number $A$ and is obtained through 
a fit to more than 200 data sets of elastic proton-nucleus scattering data 
on a wide range of nuclei that is not limited to doubly closed shell nuclei.
In comparison with electron scattering data, the DEM parametrization produces 
in general good results for doubly magic nuclei and less good but still 
acceptable results for nuclei with a number of nucleons far from the magic 
numbers \cite{esotici2,PhysRevC.89.034604}.

In Figs. \ref{fig:minervanu} and \ref{fig:minervanubar} we present the differential 
cross section $d\sigma/dQ^{2}_{QE}$ for neutrino  and antineutrino  scattering off a 
CH target as a function of the reconstructed four-momentum transfer
squared $Q^{2}_{QE}$, which is obtained, as for the experiment,  assuming an 
initial state nucleon at rest with a constant binding energy set to 34 MeV 
(30 MeV) in the neutrino (antineutrino) case. The calculated cross sections are 
then folded with the 
MINER$\nu$A neutrino and antineutrino fluxes and compared with the experimental 
data of \cite{PhysRevLett.111.022502,PhysRevLett.111.022501}. 

The RGF cross sections in Figs. \ref{fig:minervanu} and \ref{fig:minervanubar} 
are in good agreement with the data. 
Both  RGF-EDAI and RGF-DEM results are within the error bars in the entire kinematical 
range of  MINER$\nu$A. The RGF-EDAI cross sections are, however, larger than the 
RGF-DEM ones in the low four-momentum transfer squared
region, $Q^{2}_{QE}\lesssim 0.5$ GeV$^2$, while similar results are obtained 
with the two OPs for larger values of $Q^{2}_{QE}$. 
The differences between the two RGF results are due to the
different imaginary parts of the relativistic OPs adopted in
the calculations, that can give large differences in the
neutrino-nucleus cross sections at different energy and momentum transfer.
We note, however, that the differences are generally small 
and, in the case of antineutrino cross sections in Fig. \ref{fig:minervanubar},
they are always less than $10 \%$. 

A larger sensitivity to the choice of the relativistic OP is obtained at the
MiniBooNE kinematics. The results can be found 
in \cite{Meucci:2011vd,Meucci:ant}, where it is also shown that the RGF 
calculations are, in general,  in satisfactory agreement with the MiniBooNE 
cross sections.

An example of the comparison with the MiniBooNE data is presented in 
 Fig. \ref{fig:miniboone_075}, where the RGF CCQE double-differential
neutrino (antineutrino) cross sections averaged over the MiniBooNE fluxes
are displayed as a function of the muon kinetic energy $T_{\mu}$ for the 
 $\cos\vartheta_{\mu} = 0.75$ angular bin. The RGF-EDAI results have been 
 already published in \cite{Meucci:2011vd,Meucci:ant} and are shown here for 
 completeness. The RGF-DEM cross sections are only a bit larger but in general 
 close to the results obtained in \cite{Meucci:2011vd,Meucci:ant} with the 
 EDAD1 optical potential, which corresponds to a less recent energy-dependent 
 and  A-dependent parametrization \cite{Cooper:1993nx}. 
Both RGF results in Fig. \ref{fig:miniboone_075}
are in reasonable agreement with the data around the peak region, while the 
data are slightly underpredicted for large  $T_{\mu}$. In contrast, 
other models based on the IA underestimate the MiniBooNE cross 
sections and suggest that non-QE processes induced by
two-body currents can play an important role at MiniBooNE kinematics.
Our results in Figs. \ref{fig:minervanu},  \ref{fig:minervanubar}, and
\ref{fig:miniboone_075} show the same qualitative behavior and in general a
satisfactory agreement in comparison with both MiniBooNE and 
MINER$\nu$A data.

The very recent analysis in \cite{minerva-juan}, which makes use of the 
SuperScaling Approximation (SuSA) \cite{Amaro:2004bs}  and of the RMF approach
to neutrino scattering, shows that these two models, which both
underestimate the MiniBooNE data if the value $M_A = 1.03$ GeV/$c^2$
is adopted in the calculations, provide, with the same value of $M_A$, a good 
description of the MINER$\nu$A data. This is an indication that there is no 
need to enlarge the axial mass or to invoke any significant contribution 
from 2p-2h meson exchange currents and other effects beyond the IA to 
reproduce the MINER$\nu$A data. 
Our RGF cross sections at MiniBooNE kinematics in \cite{Meucci:2011vd,Meucci:ant} are
significantly larger than the RMF and SuSA ones and in better agreement with the
MiniBooNE CCQE data. The differences between our RGF and the RMF and
SuSA results are reduced at MINER$\nu$A kinematics. This is an indication that
in this kinematic situation the relevance of the inelastic contributions 
included in the RGF is reduced.
The RGF cross sections Figs. \ref{fig:minervanu} and  \ref{fig:minervanubar} 
are, however, still somewhat larger than the RMF and SuSA ones of  
\cite{minerva-juan} but in agreement with the data within the experimental errors.

In Table \ref{tab:tot} we report the values of the total neutrino and antineutrino 
cross sections per nucleon flux-averaged over the experimental fluxes from 
$1.5$ to $10$ GeV.
The results corresponding to the two RGF calculations well reproduce 
the experimental result in the case of neutrino scattering. In the case of 
antineutrino scattering, the RGF results are a bit larger than the measured 
cross section but in agreement with the experimental value within one standard 
deviation in the case of RGF-DEM and within two standard deviations in the case 
of RGF-EDAI. The SuSA and RMF results of  \cite{minerva-juan} are also shown 
for a comparison: they are smaller than the RGF results and in good agreement 
with the data.

\begin{table}[h]
\begin{center}
\caption{\small Results for the flux integrated total 
neutrino and antineutrino CCQE 
cross section per nucleon compared with the data from MINER$\nu$A
 \cite{PhysRevLett.111.022502,PhysRevLett.111.022501}. 
}
\begin{ruledtabular}
\begin{tabular}{cc}
 Neutrino   &    $\sigma$ [$10^{-38}$cm$^2$/neutron]   \\
\hline
          RGF-EDAI  &  $0.97$    \\
          RGF-DEM &    $0.91$  \\
          RMF \cite{minerva-juan}  &  $0.901$    \\
          SuSA \cite{minerva-juan} &    $0.828$  \\
          Experimental \cite{PhysRevLett.111.022502} &   
           0.93 $\pm$ 0.01 (stat) $\pm$ 0.11 (syst) \\
\hline
\hline
 Antineutrino   &    $\sigma$ [$10^{-38}$cm$^2$/proton]   \\
\hline
          RGF-EDAI  &    $0.71$   \\
          RGF-DEM &    $0.68$   \\
          RMF \cite{minerva-juan}  &  $0.583$    \\
          SuSA \cite{minerva-juan} &    $0.550$  \\
          Experimental \cite{PhysRevLett.111.022501} &  
           0.604 $\pm$ 0.008 (stat) $\pm$ 0.075 (syst) \\
\end{tabular}
\end{ruledtabular}
\label{tab:tot}
\end{center} 
\vspace*{-0.3 cm}
\end{table}

\section{Conclusions}

In this paper we have compared the predictions of the RGF model with the
CCQE neutrino and antineutrino-nucleus scattering  MINER$\nu$A data. 
The RGF model is able to give a satisfactory description of electron 
scattering cross sections in the QE region and also of the 
CCQE MiniBooNE data without the need to increase the standard value of the 
axial mass. We have shown that the RGF results obtained with the standard 
value of the axial mass are also able to describe MINER$\nu$A data for CCQE neutrino and 
antineutrino scattering. 

The RGF results are usually larger than the results of other models based on 
the IA. The differences depend on kinematics. 
The RGF model is based on the use of a complex energy-dependent relativistic 
OP whose imaginary part includes the overall effect of the inelastic 
channels which give different contributions at different energies.
The energy dependence of the OP makes the RGF results sensitive to the 
kinematic conditions of the calculations. 
With the use of a phenomenological complex OP the model includes all the allowed 
final-state channels and not only direct one-nucleon emission processes.
The important role of contributions other than direct 
one-nucleon emission has been confirmed by different independent models in the case of
MiniBooNE cross sections \cite{Benhar:2010nx,Benhar:2011wy}, but  the same  
conclusion is doubtful in the case of MINER$\nu$A data  \cite{minerva-juan}.

The RGF model does not include two-body meson exchange currents, but it can 
include rescattering processes of the nucleon in its way out of the nucleus,  
non-nucleonic $\Delta$ excitations, which may arise during nucleon propagation, 
with or without real pion production, and also some multinucleon processes. Such 
contributions are not incorporated explicitely in the model, but can be 
recovered, to some extent,  
by the imaginary part of the relativistic OP. 
The use of  a phenomenological OP, however, does not
allow us to disentangle and evaluate the role of a specific reaction process. 
For instance, we cannot disentangle the contribution of
some pion emission processes which can be taken into account by the imaginary part 
of the OP but which have been subtracted in the analyses of CCQE
data.  

The RGF results are affected by some theoretical uncertainties due to the use 
of different available parametrizations of the relativistic OP. These
uncertainties depend on kinematics and on the specific situation that is
considered. A better determination 
of the OP which fulfills the dispersion relations in the whole energy region 
of interest is required to reduce the theoretical uncertainties  of the model 
and deserves further investigation.

\begin{acknowledgements}

We thank M. V. Ivanov for useful and helpful discussions in the preparation of the manuscript.

\end{acknowledgements}

%

\end{document}